# Promoting Component Reuse by Separating Transmission Policy from Implementation

Scott M. Walker
*University of St Andrews*
*scott@dcs.st-and.ac.uk*

Alan Dearle
*University of St Andrews*
*al@dcs.st-and.ac.uk*

Graham N. C. Kirby
*University of St Andrews*
*graham@dcs.st-and.ac.uk*

Stuart Norcross
*University of St Andrews*
*stuart@dcs.st-and.ac.uk*

## Abstract

*In this paper we present a methodology and set of tools which assist the construction of applications from components, by separating the issues of transmission policy from component definition and implementation. This promotes a greater degree of software reuse than is possible using traditional middleware environments.*

*Whilst component technologies are usually presented as a mechanism for promoting reuse, reuse is often limited due to design choices that permeate component implementation. The programmer has no direct control over inter-address-space parameter passing semantics: it is fixed by the distributed application's structure, based on the remote accessibility of the components. Using traditional middleware tools and environments, the application designer may be forced to use an unnatural encoding of application level semantics since application parameter passing semantics are tightly coupled with the component deployment topology.*

*This paper describes how inter-address-space parameter passing semantics may be decided independently of component implementation. Transmission policy may be dynamically defined on a per-class, per-method or per-parameter basis.*

## Introduction

During remote method call, different transmission policies can be applied to components that are passed across address space boundaries. The requirements of each particular application dictate the parameter passing semantics applied to particular arguments and return values. Typically, components are either passed-by-reference or passed-by-value though variations such as pass-by-migrate or pass-by-visit exist. This paper describes a methodology and set of tools that allow an application programmer to separate the issues of component transmission policy from component definition and implementation. The advantage of this separation is that it aids component reusability since components can be used in applications in a more flexible manner.

An environment in which transmission policy is distinct from component implementation promotes software reuse to a greater degree than is possible using traditional middleware systems. A single component can be reused in multiple applications with different parameter passing semantics without the need to modify the component.

Consider the following use-case: Some address book software models each entry in the address book as a component. The software runs on a desktop machine and holds references to these components. Using traditional middleware the PDA must either obtain the components by-reference, meaning that they are unavailable when disconnected from the network, or by-value, meaning coherency control must be performed on each update. Using the described technology, the PDA can obtain components by-reference while it is connected to the network, obviating the need for coherency control on update. Only when disconnecting from the network does the PDA obtain components by-value in order that they remain available offline. At any given moment, the programmer can employ the most advantageous transmission policy for the circumstances.

By allowing the specification of transmission policy dynamically and independently of component implementation, the roles of component programmer and application programmer are separated. The component programmer is concerned only with the functional requirements of the components, not the parameter passing semantics that may be applied to it when it is deployed in an application. Components make fewer assumptions about the environment in which they are to be used. The application programmer has the freedom to apply any transmission policy to any component, thereby increasing the likelihood that any given component will be reusable in another context.

The technology that permits the separation of transmission policy from component creation has been implemented as part of a middleware system known as the *RAFDA Run Time (RRT)*. No special steps need be taken during component implementation and components can be assembled into applications in the conventional manner. Both component programmer and application programmer can benefit from using the RRT without having to alter their development process.

Typical middleware systems do not allow this separation of transmission policy from implementation. Transmission policy is decided statically and cannot be changed without modifying the component. This inflexibility hampers reuse. The application programmer has no control over the transmission policy applied to components – it is hard-coded by the component programmer.

Component transmission policy is commonly based on whether a component is remotely accessible. Choosing transmission policy in this manner can force the application programmer to use an unnatural encoding of application level semantics. Either the application programmer must create an application within the constraints of the available components or the component programmer must know at component creation time the semantics of the application in which it is to be deployed.

This paper describes several notable traits of the RRT middleware system, namely the following:

- A single component can be passed-by-reference or passed-by-value.
- Transmission policy and application distribution are not tightly coupled.
- Inter-address-space parameter passing semantics can be controlled.

Using the RRT, components are written without regard for their transmission policy. Applications are constructed from components in the usual manner and an application-specific component transmission policy is specified separately. This transmission policy can be dynamically altered and is defined on a per-class, per-method or per-parameter basis.

## Related work

Whilst component technologies are usually presented as a mechanism for promoting reuse, this reuse is often limited due to design choices that permeate component implementation. During the creation of a distributed application, the programmer is forced to decide statically how the application is partitioned. Particular component classes are written to be remotely accessible. The transmission policy applied to components is decided statically.

Using Java RMI[1] and Microsoft .NET remoting[2] the programmer defines special remote classes, instances of which are remotely accessible. During remote calls, instances of (almost[1]) any class can be passed as arguments. Arguments that are instances of remotely accessible classes are always passed by-reference. Arguments that are not instances of remotely accessible classes are always passed by-value. The parameter passing semantics applied to components are inflexible and tightly coupled with the distribution of the application.

Using CORBA v2.3 or later[3], the component programmer decides statically at component creation time whether a component will cross network boundaries by-reference or by-value. Initially, this seems to offer moderately more control than RMI and .NET remoting. However, only components specified as CORBA components can be passed as arguments, unlike the others which permit other classes of component to cross network boundaries, if only by-value.

Web Services[4] technologies permit only pass-by-value. The RRT includes some extensions to the Web Services model that support pass-by-reference and are described outwith this paper[5].

In all cases, typical middleware systems restrict reusability and application semantics in the following ways:

- They define transmission policy statically. A component can only be passed-by-reference or passed-by-value for the entire duration of the application.
- They tightly couple transmission policy and application distribution.
- The application programmer has no direct control over inter-address space parameter passing semantics.

These restrictions hamper component reuse because application level semantics are built into components at creation time in an unalterable fashion. We have created the RRT in order to overcome these limitations.

---

[1] Instances must be of a remotely accessible or serializable class

The RRT has several features that differentiate it from typical middleware systems, namely:

- The provision of a transmission policy framework that allows the dynamic definition of transmission policy on a:
  - Per-parameter basis
  - Per-method basis
  - Per-class basis
- If an application component can cross network boundaries then the RRT can choose whether to pass it by-reference or by-value

The RRT is capable of deploying arbitrary components as Web Services. These components can be referenced from remote address-spaces using a remote reference scheme implemented by the RRT. If a component is to be passed by-reference, the RRT will automatically deploy the component to make it remote accessible and will transmit a remote reference across the network. If the component is to be passed by-value, the RRT will serialize the component and transmit it across the network. This functionality is described in detail elsewhere.

The RRT is capable of transmitting any component by-reference and any component by-value. The RRT dynamically decides how to treat each component based on the transmission policy. To exploit this mechanism, the programmer must be able to define transmission policy in an expressive and flexible manner.

## Defining transmission policy

During remote method call, components are passed across address space boundaries as arguments and return values. Transmission policy dictates the manner in which components are encoded for transmission. It decides which parameter passing semantics will be employed during remote method calls.

Though the transmission policy framework has been described in the context of the RRT it is applicable with any middleware. The RRT supports passing parameters by-reference or by-value but the described transmission policy framework not restricted to these two mechanisms. It is scalable to accommodate any parameter passing mechanisms that the underlying middleware supports.

In order to define the transmission policy for an application, the programmer specifies a series of *policy rules*. There are three kinds of policy rule:

- Parameter policy rules
- Method policy rules
- Class policy rules

*Parameter policy rules* are associated with individual method parameters. They indicate how particular method arguments should be passed across address-space boundaries during a call to the specified method. They allow fine-grained control over the transmission policy that is applied to the parameters of a method. For example, a parameter policy rule might specify that during a call to a particular method, the second parameter should be passed-by-value.

*Method policy rules* are associated with methods as a whole. They have a dual role. They specify how return values from methods should be passed across address-space boundaries. For example, a method policy rule might specify that during a call to a particular method, the return value should be passed-by-reference. Additionally, they allow a single transmission policy to be associated with all parameters of a method, avoided the need to specify a parameter policy rule for each. For example, a method policy rule might specify that during a call to a particular method, all parameters should be passed-by-value.

*Class policy rules* are associated with classes. They indicate how instances of classes should be passed across address-space boundaries. For example, a class policy rule might specify that all instances of a particular class are passed-by-value. Each class policy rule applies to exactly one class. It does not apply to sub-classes of that class. Class policy rules are applied based on the actual classes of the parameters, not the those specified in the method signature.

Policy rules apply only in the address space in which they are specified, though they apply to all components in that address space. Policy rules can be specified dynamically at any point during application execution and they come into force immediately. A component programmer can effectively specify policy rules statically by specifying them in the component's initialization code. For example, in Java, policy rules specified in the static initializer are active from class load time.

This functionality distinguishes the RRT from typical middleware systems and tackles the limitations listed at the beginning of this section. The dynamic specification of policy rules that dictate application parameter passing semantics returns control of these semantics to the application programmer.

Policy rules are created through the *policy manager*. There is a single policy manager per address-space which is responsible for evaluating transmission policy in that address-space. Each policy manager stores a database of policy rules, specified by the application programmer.

Transmission policy is concerned with cross-address-space communication and so is applied at serialization time. During serialization, the policy manager determines the transmission policy that should be applied to each component it is asked to serialize. When a component is serialized by-value, the components it references are also passed into the serializer. The policy manager will determine how each of these components should be passed across the network and serialize them appropriately. Method and parameter policy rules can be specified with a depth value. This depth indicates how far into the closure of an argument the policy rule applies.

The policy manager provides the methods shown in Figure 1 for the specification of policy rules. The *Policy* class is not shown. It is an enumeration class identifying all available parameter passing mechanisms. The purpose of the *isOverridable* flag is discussed later.

```
public static void setClassPolicy(
          String className,
          Policy policy,
          boolean isOverridable
          ) {...}
public static void setMethodPolicy(
          String className,
          String methodName,
          Policy policy,
          boolean isOverridable
          ) {...}
public static void setParamPolicy(
          String className,
          String methodName,
          int paramNumber,
          Policy policy,
          boolean isOverridable
          ) {...}
```

**Figure 1: Policy manager methods used to specify policy rules**

## Evaluating transmission policy

The policy manager makes all transmission policy decisions for all components in its address space. However, from the perspective of a single component, the programmer may wish to control the transmission policy that is applied to the following:

1. The arguments the component passes when calling some remote method
2. The arguments the component receives when a method is called on it
3. The return value the component transmits after a method has been called on it
4. The return value the component receives after calling some remote method

One component's passed arguments are another's received arguments. For some remote method call, the caller may wish to apply one transmission policy (case 1 above) while the callee wishes to apply another transmission policy (case 2 above). Cases 3 and 4 exhibit the same problem.

Each policy manager has direct control over the transmission policy applied to components outgoing from its address space, that is, cases 1 and 3 above. They cannot have direct control over components that are incoming from a remote address space, that is, cases 2 and 4 above. When evaluating transmission policy during a remote call, a policy manager may solicit information from the policy manager in the remote address space about the policy rules it has associated with this remote call.

Individual policy managers are configured to either use this information from the remote policy manager or to base the transmission policy decision on locally specified rules alone. A policy manager that considers the remote policy manager is known as a *co-operative* policy manager. All policy managers, whether co-operative or not, respond to requests for information about their locally specified policy rules.

From the specified policy rules, the transmission policy applicable to a particular remote method call can be deduced. It is based on the class of the component; the method being called; whether the component is an argument or return value; and the depth of the component in the argument's closure. Figure 2 shows the methods provided by the policy manager that determine transmission policy.

```
public static TransmissionPolicy
     getTransmissionPolicy(
          String className,
          String methodName,
          int paramNumber,
          Object param,
          int depth) {...}
public static TransmissionPolicy
     getReturnTransmissionPolicy(
          String className,
          String methodName,
          Object returnValue,
          int depth) {...}
```

**Figure 2: The policy manager methods used to evaluate transmission policy**

These methods are called by the RRT during component serialization or by a co-operative remote policy manager that is evaluating transmission policy. In addition to specifying how the component should be passed across the network, the returned *TransmissionPolicy* also contains information about the kind of policy rule that was used to make the decision. This information is used by remote policy managers but is ignored by the RRT.

## Resolving policy rule contention

Clearly, there is scope for contention between policy rules specified in different policy managers. A class policy rule in one address-space can specify that instances of X are always passed-by-value, while a class policy rule in another address-space specifies that instances of X are always passed-by-reference.

Similarly, contention can exist among rules specified within a single address-space. For example, a component of class X is passed as a parameter to method m(). A class policy rule may indicate that instances of X are passed-by-value while a method policy rule simultaneously indicates that parameters to method m() are always passed-by-reference.

The policy manager has a set of policy rules, including some that may have been received from a remote policy manager, and must decide which to apply.

A hierarchy of policy rules is defined. Higher rules are followed while lower rules are ignored. The hierarchy is:

1. Parameter policy rule
2. Method policy rule
3. Class policy rule
4. Default policy

A parameter policy rule is followed before all others. If none exists, then the policy manager looks for an applicable method policy rule. If none exists, then it looks for an applicable class policy. If no policy rules have been defined then a default policy is applied. The policy rule that is used to decide the transmission policy in a particular set of circumstances is known as the *dominant* rule.

This strict hierarchy is restrictive. Under some circumstances, it is desirable that a class policy rule take precedence over a parameter or method policy rule. For this reason, policy rules are specified with a flag that indicates whether the rule can be overridden. A rule that cannot be overridden is always followed before a rule that can be overridden, irrespective of their hierarchical position. The hierarchy can be revised as follow:

1. Parameter policy rule (non-overridable)
2. Method policy rule (non-overridable)
3. Class policy rule (non-overridable)
4. Parameter policy rule (overridable)
5. Method policy rule (overridable)
6. Class policy rule (overridable)
7. Default policy

It is recommended that policy rules are specified as overridable in most circumstances. Despite specifying a class policy rule as non-overridable, it will still be overridden by a non-overridable method policy. The authors suggest that it should rarely be necessary to override a policy rule that has been specified as non-overridable and that such an operation should be performed with care.

The policy manager holds a series of policy rules that are applicable during a particular method call. The transmission policy received from a remote policy manager also includes the dominant rule in that remote address-space. The hierarchy can resolve contention among this set of policy rules.

Contention can still exist occur if the dominant rule in the remote policy manager is hierarchically equivalent to the dominant rule in the local policy manager. Contention of this form is resolved differently depending on whether the transmission policy is associated with an argument of a return value.

The callee's policy rule is followed over the caller's when choosing the transmission policy for arguments. Conversely, the caller's policy rule is followed over the callee's when choosing the transmission policy for the return value. The RRT is capable of deserializing and using components irrespective of the transmission policy used during serialization. The programmer is responsible for ensuring that application transmission policy is specified in a consistent manner that leads to the desired application semantics.

## Future Work

Initial measurements indicate that the cost of dynamically evaluating transmission policy is subsumed by the cost of serialization leading us to believe that the benefits gained outweigh the expense. We intend to perform further measurements to evaluate the trade-off in more detail.

We hope to introduce additional features to the transmission policy framework. Currently, policy managers hold policy rules that apply only to components in the local address space. Policy managers can co-operate with each other in order to reach a consensus but the specification of policy rules that apply across the entire application is not supported. We propose the introduction of a mechanism that peers together policy managers such that policy rules defined in any one of them apply in all. It would be possible to peer together only a subset of the policy managers active in a distributed system while the remainder stay autonomous.

We propose extensions to the policy rules. It will be possible to specify a class policy rule that applies not just to a single class, but to the class's entire inheritance hierarchy. The programmer will be also able to specify policy rules that apply only to a single call.

## Conclusion

This paper describes the transmission policy framework provided by the RAFDA Run Time (RRT). The RRT overcomes the limitations inherent in typical middleware systems with respect to component transmission policy and subsequently, their reuse. The RRT separates the specification of the parameter passing semantics applied to a component during inter-address-space method call from the component's creation and implementation. It provides a mechanism and framework to allow the dynamic specification of component transmission policy on a per-class, per-method or per-parameter basis.

Application semantics are no longer driven by decisions made statically during component creation. This aids component reuse since the programmer has complete control over application semantics independently of the component implementation.